\documentclass[aps,prd,showpacs,preprintnumbers,nofootinbib,twocolumn]{revtex4}
\usepackage[dvips]{graphicx}
\usepackage{bm,latexsym,amsmath,amssymb,amsfonts,mathrsfs}
\newcommand*{\D}{{\rm d}}
\newcommand*{\omm}{\Omega_{\rm m}}
\newcommand*{\mpl}{M_{\rm Pl}}
\begin{document}

\title{Cosmic expansion and growth histories in Galileon scalar-tensor models of dark energy}

\author{Tsutomu~Kobayashi}
\email[Email: ]{tsutomu"at"resceu.s.u-tokyo.ac.jp}
\affiliation{Department of Physics, Waseda University, Okubo 3-4-1, Shinjuku, Tokyo 169-8555,
Japan\footnote{Present address: Research
Center for the Early Universe (RESCEU), Graduate School of Science, The
University of Tokyo, Tokyo 113-0033, Japan }}

\begin{abstract}
We study models of late-time
cosmic acceleration in terms of scalar-tensor theories generalized to include
a certain class of non-linear derivative interaction of the scalar field.
The non-linear effect suppress the scalar-mediated force at short distances
to pass solar-system tests of gravity.
It is found that the expansion history until today is almost indistinguishable from
that of the $\Lambda$CDM model or some (phantom) dark energy models,
but the fate of the universe depends clearly on the model parameter.
The growth index of matter density perturbations is computed to show that
its past asymptotic value is given by 9/16, while the value today is as small as 0.4.
\end{abstract}

\pacs{98.80.-k, 04.50.Kd}
\preprint{WU-AP/307/10}
\maketitle

\section{Introduction}

Since the discovery of cosmic acceleration
various possibilities have been explored to account for this mystery.
Probably the most conservative possibility is that
the accelerated expansion arises from a cosmological constant
or some unknown dynamical field (e.g., a quintessence)~\cite{Copeland:2006wr}.
Instead of introducing a new component in ``$T_{\mu\nu }$''
one may alternatively consider that
dark energy is {\em geometric}, i.e.,
modification to general relativity (GR) at long distances
is responsible for cosmic acceleration. 
The latter class of models includes
scalar-tensor theories, $f(R)$ gravity~\cite{f(R)},
and the Dvali-Gabadadze-Porrati (DGP) braneworld~\cite{DGP, DDG}.

A difficulty in modifying GR at long distances lies, however, on
short distance scales;
a new scalar gravitational degree of freedom,
which commonly appears in geometric dark energy models,
must be tamed carefully in order to pass the stringent tests of gravity in the solar system~\cite{will, uzan-review}.
Two main ways of hiding the scalar degree of freedom are known to exist.
The first one is to make the scalar effectively massive in the vicinity of matter.
This is called the chameleon mechanism~\cite{chameleon}
and is utilized in viable $f(R)$ models~\cite{viablefr}.
The second way is decoupling the scalar from matter in the vicinity of the matter sources.
This is known as the Vainshtein mechanism~\cite{Vs},
and the DGP braneworld offers a nice example that implements it
by a non-linear self-interaction in the kinetic term.
See also a recent paper~\cite{symmetron} for yet another
way of suppressing the scalar-mediated force.

Inspired by the DGP braneworld,
a class of scalar-tensor theories of gravity has been explored
that enjoys self-screening of the scalar-mediated force in the vicinity of matter
due to non-linear derivative interactions.
Such a scalar degree of freedom is called the Galileon field
in the original proposal~\cite{G1}, because
the equation of motion for the scalar $\phi$
is invariant under $\partial_\mu\phi\to\partial_\mu\phi+b_\mu$
on Minkowski spacetime.
The Galileon scalar-tensor theories have been covariantized and
a general form of the Lagrangian in curved spacetime
has been considered in~\cite{G2}\footnote{Upon covariantization
the original symmetry under $\partial_\mu\phi\to\partial_\mu\phi+b_\mu$ is lost.
However, covariant Galileon Lagrangians are uniquely determined
by requiring that the equations of motion have only second derivatives of the fields.
Thus, the name ``Galileon'' may be misleading and inappropriate. Nevertheless,
we use the name in this paper because no other name is as good as ``Galileon.''}
(see also~\cite{GE1, GE2}).
The resultant equations of motion have only second derivatives of the fields,
and hence the Galileon theories are in some sense similar to Lovelock gravity~\cite{Lovelock}.
Cosmology based on a Galileon field has been studied in Refs.~\cite{CK, SK, KTS}.

In this paper, we study aspects of Galileon cosmology in terms of
a certain wider class of the Lagrangian than previously studied.
The action we consider is of the form
\begin{eqnarray}
S&=&\int\D^4x\sqrt{-g}\biggl[
\phi R -\frac{\omega(\phi)}{\phi}(\nabla\phi)^2
\cr&&\qquad\qquad\qquad
+\frac{\lambda^2(\phi)}{\phi^2}\Box\phi(\nabla\phi)^2
+{\cal L}_{\rm m}
\biggr],\label{action}
\end{eqnarray}
where $\phi$ is the Galileon field and
the coupling $\lambda(\phi)$ has dimension of length.
Matter (represented by the Lagrangian ${\cal L}_{\rm m}$)
is universally coupled to gravity in the Jordan frame.
The gravitational field equations derived from~(\ref{action}) are
\begin{eqnarray}
\phi G_{\mu\nu}&=&\frac{1}{2}T_{\mu\nu}+\nabla_\mu\nabla_\nu\phi-g_{\mu\nu}\Box\phi
+\frac{\omega}{\phi}\biggl[\nabla_\mu\phi\nabla_\nu\phi
\cr&&
-\frac{1}{2}g_{\mu\nu}(\nabla\phi)^2\biggr]
-\frac{1}{2}g_{\mu\nu}\nabla_\lambda\left[\frac{\lambda^2}{\phi^2}(\nabla\phi)^2\right]\nabla^\lambda\phi
\cr&&
+\nabla_{(\mu}\left[\frac{\lambda^2}{\phi^2}(\nabla\phi)^2\right]\nabla_{\nu)}\phi
-\frac{\lambda^2}{\phi^2}\nabla_\mu\phi\nabla_\nu\phi\Box\phi,
\end{eqnarray}
and the equation of motion for the Galileon field is given by
\begin{eqnarray}
&&R+\omega\left[\frac{2\Box\phi}{\phi}-\frac{(\nabla\phi)^2}{\phi^2}\right]
+\frac{\omega'}{\phi}(\nabla\phi)^2
\cr&&\quad
+\frac{2\lambda^2}{\phi^2} \left[\nabla_\mu\nabla_\nu\phi\nabla^\mu\nabla^\nu\phi
-(\Box\phi)^2+R_{\mu\nu}\nabla^\mu\phi\nabla^\nu\phi \right]
\cr&&\quad
+4\left(\frac{\lambda^2}{\phi^2}\right)'\nabla_\mu\phi\nabla_\nu\phi\nabla^\mu\nabla^\nu\phi
+\left(\frac{\lambda^2}{\phi^2}\right)''(\nabla\phi)^2(\nabla\phi)^2
\cr&&\quad
=0,
\end{eqnarray}
where
a prime denotes differentiation with respect to $\phi$.
The above modified gravity theory may be regarded as the generalization of Brans-Dicke gravity, but
the cubic derivative interaction, $[\lambda^2/\phi^2]\Box\phi(\nabla\phi)^2$,
plays a key role in manifesting the Vainshtein mechanism.
The Vainshtein radius, below which
the Galileon-mediated force is screened,
is evaluated as
$r_V\sim [r_g\lambda^2/(1+2\omega/3)^2]^{1/3}$,
where $r_g$ is the Schwarzschild radius of the matter source.

In what follows we shall consider
the particular case with $\omega=$ const and $\lambda^2\propto \phi^\alpha$,
where $\alpha$ may be positive and negative.
The self-accelerating de Sitter universe of~\cite{SK}
corresponds to the subclass $\alpha=0$.
We are going to study the modified gravity theory with general $\alpha$
as a possible alternative to a cosmological constant or dynamical dark energy.
Accelerating cosmologies from a scalar field non-minimally coupled to gravity
{\em without} the non-linear derivative interaction have been studied extensively
in~\cite{extquint, Reconst}.

Since the background expansion history in successful modified gravity
is by definition almost identical to that of the standard $\Lambda$CDM model or
other dynamical dark energy models,
it is important to study the growth history of perturbations
as a tool to distinguish modified gravity from models with a cosmological constant or dark energy.
We compute the growth index of matter
density perturbations~\cite{Peeb,WS} in Galileon cosmology,
which is known to be a powerful discriminant among models of
cosmic acceleration~\cite{Linder1,Linder2, P-G}.

The paper is organized as follows.
In the next section we study the background evolution of Galileon cosmology
derived from the above action. Then, in Sec.~III, we discuss the growth density perturbations
in Galileon cosmology, paying particular attention to the growth index.
The final section is devoted to conclusions.

\section{Cosmic expansion in Galileon theories}

Let us start with investigating the background evolution of Galileon cosmology
in the presence of cold dark matter.
It is convenient to define a new field by $\phi = (\mpl^2/2)\exp (\sigma)$.
For a flat Friedmann-Robertson-Walker universe,
$\D s^2 = -\D t^2+a^2(t)\D\mathbf{x}^2$,
the gravitational field equations are
\begin{eqnarray}
3H^2&=&\frac{\rho_{\rm m}}{\mpl^2}e^{-\sigma}
-3H\dot\sigma+\frac{\omega}{2}\dot\sigma^2
+3\lambda^2 H\dot\sigma^3
\nonumber\\&&
-\frac{1}{2}\left[(\lambda^2)_{,\sigma}-2\lambda^2\right]\dot\sigma^4,
\label{bg_Friedmann}
\\
-3H^2-2\dot H&=&\ddot\sigma+\dot\sigma^2+2H\dot\sigma+\frac{\omega}{2}\dot\sigma^2
\nonumber\\&&
-\lambda^2\dot\sigma^2\ddot\sigma-\frac{(\lambda^2)_{,\sigma}}{2}\dot\sigma^4,
\label{bg_ij}
\end{eqnarray}
and the Galileon field equation is
\begin{eqnarray}
&&6\left(2H^2+\dot H\right)-\omega\left(2\ddot\sigma+\dot\sigma^2+6H\dot\sigma\right)
-\omega_{,\sigma}\dot\sigma^2
\nonumber\\&&\quad
-6\lambda^2\left(3H^2\dot\sigma+\dot H\dot\sigma+2H\ddot\sigma+2H\dot\sigma^2\right)\dot\sigma
\nonumber\\&&\quad
+4\left[(\lambda^2)_{,\sigma}-2\lambda^2\right]\dot\sigma^2\ddot\sigma
+\left[(\lambda^2)_{,\sigma\sigma}-(\lambda^2)_{,\sigma}-2\lambda^2\right]\dot\sigma^4
\nonumber\\&&\quad
=0.
\label{bg_Galileon}
\end{eqnarray}
Here,
an overdot represents differentiation with respect to $t$ and
$H(t):=\dot a/a$ is the Hubble expansion rate.
The energy density of cold dark matter, $\rho_{\rm m}$, is conserved in the Jordan frame,
$\dot{\rho}_{\rm m}+3H\rho_{\rm m}=0$.
The Friedmann equation~(\ref{bg_Friedmann})
can be written in the form
\begin{eqnarray}
3 H^2 =\frac{1}{\mpl^2}\left(\rho_{\rm m} + \rho_{\rm eff}\right),
\end{eqnarray}
where the effective dark energy density $\rho_{\rm eff}$ is defined as
\begin{eqnarray}
\frac{\rho_{\rm eff}}{\mpl^2}&=&e^\sigma\biggl[
-3H\dot\sigma+\frac{\omega}{2}\dot\sigma^2
+3\lambda^2 H\dot\sigma^3
\nonumber\\&&
-\frac{1}{2}\left(\lambda^2_{,\sigma}-2\lambda^2\right)\dot\sigma^4\biggr]
+3H^2(1-e^\sigma).
\end{eqnarray}

We assume that $H_0^{-2}, \lambda_{,\sigma}^2,\lambda^2_{,\sigma\sigma}\lesssim{\cal O}(\lambda^2)$,
where $H_0$ is the Hubble rate at present. At early times, $H^{-2}\ll \lambda^2$,
Eq.~(\ref{bg_Galileon}) is approximately solved to give
\begin{eqnarray}
\lambda^2  \dot\sigma^2  \approx \frac{2H^2+\dot H}{3H^2+\dot H}\approx\frac{1}{3},
\label{early-time-sol}
\end{eqnarray}
where we used $H\approx 2/3t$.
(To be precise, this is the consequence of $\rho_{\rm m}\gg{\cal O}(\rho_{\rm eff})$,
which will be verified shortly.) 
From this we see that
if the variation of $\lambda$ is small over the Hubble time scale,
$\dot\sigma$ is constant and $\sigma\approx\dot\sigma t\approx2\dot\sigma/3H$.
Note that with the rescaling of $\mpl$ we may set $\sigma(t=0)=0$.
Two branches of solutions are present, $\dot\sigma>0$ and $\dot\sigma<0$,
but it turns out that
the negative $\dot\sigma$ branch is plagued with ghost instabilities and
only the positive $\dot\sigma$ branch is healthy~\cite{CK} (see the Appendix).
We therefore focus on the case with $\dot\sigma>0$ in this paper.
In this early-time regime, we have
$\rho_{\rm eff}/\mpl^2\approx -2H\dot\sigma e^\sigma+3H^2(1-e^\sigma)\approx-4H\dot\sigma$,
so that the usual matter-dominated universe is indeed consistently reproduced,
$3\mpl^2H^2\simeq \rho_{\rm m}\gg{\cal O}(\rho_{\rm eff})$.
This is the cosmological version of the Vainshtein effect~\cite{CK}.
It is worth noting that for $\dot\sigma>0$
the effective dark energy density is negative in the initial stage.
The behavior $\rho_{\rm eff}\approx-4H\dot\sigma\sim - H/\lambda$
reminds us of the modified Friedmann equation in the normal branch of the DGP model~\cite{DDG}.

We then consider the late-time evolution of Galileon cosmology.
To be more specific, we study the case with $\omega=$ const and
\begin{eqnarray}
\lambda^2 = r_c^2\left(\frac{2\phi}{\mpl^2}\right)^\alpha=r_c^2e^{\alpha\sigma}.\label{def:coupling}
\end{eqnarray}
The model with $\omega=0$ and $\alpha=-1$ was studied in Ref.~\cite{CK}.
It was shown in Ref.~\cite{SK} that
the model with $\omega < -4/3$ and $\alpha=0$ allows for
a late-time de Sitter solution, mimicking the $\Lambda$CDM model.
We generalize the result of~\cite{SK} and show that
the more general class of models described by~(\ref{def:coupling})
can give rise to the cosmic expansion history similar to the $\Lambda$CDM model.

For general $\alpha$ one finds that
Eq.~(\ref{bg_ij}) admits the following solution:
\begin{eqnarray}
\frac{\dot H}{H^2} = -\frac{1}{2}\alpha s,\quad
\frac{\dot\sigma}{H} =s,
\label{asym-1}
\end{eqnarray}
where
\begin{eqnarray}
s:=\frac{-(2-\alpha)\pm\sqrt{-6\omega-(2-\alpha)(4+\alpha)}}{\omega+2-\alpha}.\label{asym-2}
\end{eqnarray}
In order for $s$ to be real, we require $\omega < -(2-\alpha)(4+\alpha)/6$.
This includes the parameter range in which
the scalar field would be a ghost
in the usual Brans-Dicke theory, i.e., $\omega<-3/2$, but
the situation is drastically changed by the introduction of the non-linear derivative interaction
and the model does not suffer from the pathology (see the Appendix).
For negatively large $\omega$, we have $s\sim{\cal O}(|\omega|^{-1/2})\ll 1$ and hence
$|\dot H/H^2|\ll 1$, leading to a quasi-de-Sitter expansion.
The Hubble rate is explicitly given by $H=\left(\alpha st +C\right)^{-1}$,
where $C$ is an integration constant. Equation~(\ref{bg_Friedmann}) with $\rho_{\rm m}=0$ gives
$\lambda^2(\sigma)H^2 = (2+s)/s^3$.
The scale $r_c$ must be tuned to satisfy this relation.

\begin{figure}[tb]
  \begin{center}
    \includegraphics[keepaspectratio=true,height=55mm]{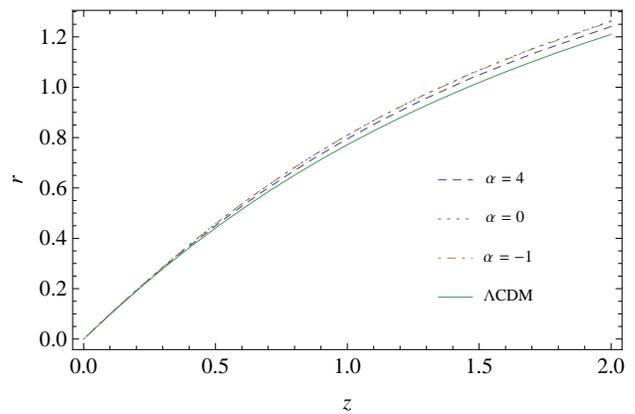}
  \end{center}
  \caption{Dimensionless physical distance for $\omega=-500$ and various $\alpha$.}%
  \label{fig:distance.eps}
\end{figure}
\begin{figure}[tb]
  \begin{center}
    \includegraphics[keepaspectratio=true,height=55mm]{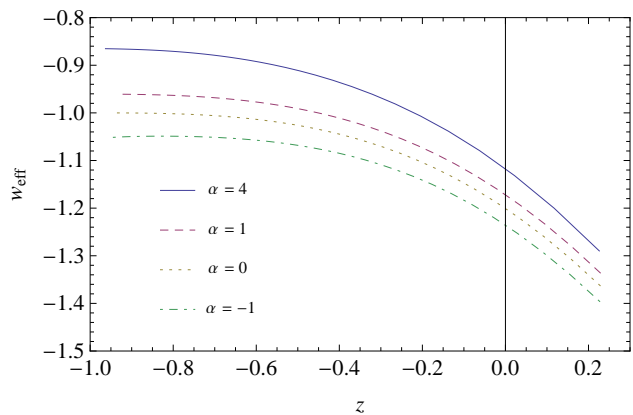}
  \end{center}
  \caption{Effective equation of state $w_{\rm eff}$ as a function of $z$ for
  $\omega=-500$ and various $\alpha$.}%
  \label{fig:weff.eps}
\end{figure}

The solution~(\ref{asym-1}) is our candidate for the future asymptotic solution.
In order to work out the background evolution
from the beginning of the matter-dominated stage
and to see the future asymptotic behavior of the cosmological solution,
Eq.~(\ref{bg_ij}) and~(\ref{bg_Galileon}) are solved numerically
with the initial conditions given by the above early-time solution.
In Fig.~\ref{fig:distance.eps} we show the dimensionless physical distance,
$r(z):=H_0 \int^z_0\D z'/H(z')$, for $\omega=-500$ and various $\alpha$.
Here, $z$ is the redshift and the present time $t_0$ is determined
from $\rho_{\rm m}/3\mpl^2H^2|_{t=t_0}=\Omega_{{\rm m}0}$,
where we use $\Omega_{{\rm m}0}=0.3$ throughout the paper.
We see that the cosmic expansion histories in the Galileon models
mimic that of the $\Lambda$CDM model,
and the difference is less than 5 percent for the plotted examples.
A closer look shows that Galileon cosmology is more similar to
phantom dark energy models at the background level,
as is obvious also from the behavior of the effective equation of state $w_{\rm eff}$ that will be
investigated shortly.
One can confirm that
the initial deviation of $\dot\sigma$
from the value given by Eq.~(\ref{early-time-sol})
does not change the result.

The effective equation of state, $w_{\rm eff}:=-\dot\rho_{\rm eff}/3H\rho_{\rm eff}-1$,
is shown in Fig.~\ref{fig:weff.eps} for $\omega=-500$ and various $\alpha$.
We plot the range $t\gtrsim t_0$ so that
we can see the asymptotic behavior $1+w_{\rm eff}\to -2\dot H/3H^2 \approx \alpha s/3$.
A phantom-like behavior is found at low $z$,
but $1+w_{\rm eff}$ gets positive eventually for $\alpha>0$, crossing the phantom divide.
Therefore, the phantom-like behavior is temporary and the big rip singularity can be avoided in this case.
In contrast, $1+w_{\rm eff}$ remains negative in the future for $\alpha<0$,
leading to the singular fate of the universe, $H\to \infty$.

Since $\rho_{\rm eff}\propto H\approx 2/3t$ at early times, we have
$w_{\rm eff}\to-1/2$ as $z\to\infty$.
The effective dark energy density is negative initially
and evolves into $\rho_{\rm eff}>0$ at lower $z$,
which implies that $\rho_{\rm eff}$ crosses zero at some $z=z_*$ and
we have $w_{\rm eff}\to\pm \infty$ as $z\to z_*^\pm$.
However, this divergence is an artifact of the definitions of $\rho_{\rm eff}$ and $w_{\rm eff}$
and no anomaly is seen in the evolution of $H(z)$.
The whole behavior of $w_{\rm eff}$ is most similar to
what is found in the normal branch of the DGP braneworld
with the additional cosmological constant or a quintessence
field on the brane~\cite{SS, ndgp1, ndgp2}.

\section{Density perturbations}

\subsection{Some preliminaries}

In various dark energy models and modified gravity models
the governing equations
for the homogeneous background and the cold dark matter density perturbation $\delta$ can often be
written in the form
\begin{eqnarray}
&&3\mpl^2H^2 = \rho_{\rm m}+\rho_{\rm eff},\label{bg_gen}
\\
&&\ddot \delta+2H\dot\delta = \frac{\rho_{\rm m}}{2\mpl^2}\xi\delta.\label{delta_gen}
\end{eqnarray}
Here, $\xi=\xi(t)$ represents the ratio of the effective gravitational coupling ``$G_{\rm eff}$''
to $G\,[:=1/(8\pi \mpl^2)]$,
which in general differs from unity.\footnote{The
effective gravitational coupling may depend on the scale $k$,
but we do not care about this point as we are not interested in
the power spectrum of $\delta$ in this paper.}
We shall show in the next subsection that the perturbation equation
indeed takes the form of~(\ref{delta_gen}) in Galileon theories.
Before going to the analysis of the specific cases of Galileon modified gravity,
we explain how we can characterize the growth of density perturbations
that follows from~Eqs.~(\ref{bg_gen}) and~(\ref{delta_gen}).

We are going to study
the growth index $\gamma$ of density fluctuations.
The growth index is defined via the growth rate
$f=\D\ln\delta/\D\ln a$ as
\begin{eqnarray}
\gamma =\frac{\ln f}{\ln\Omega_{\rm m}(a)},
\end{eqnarray}
where $\Omega_{\rm m}(a) = \rho_{\rm m}/(3\mpl^2 H^2)$.
This quantity was pioneered by~\cite{Peeb, WS},
and it turned out that
$\gamma$ is a powerful and useful discriminant among different models of cosmic acceleration
as $\gamma$ has a different but almost constant value depending on the model~\cite{Linder1, Linder2, P-G}.
For instance, $\gamma \simeq 0.55$ for the $\Lambda$CDM model and $\gamma \simeq 0.68$
for the DGP model.
The purpose of this subsection is to rederive the useful formula for the asymptotic value
of $\gamma$, $\gamma(z\to\infty)=:\gamma_\infty$~\cite{Linder2}.

In terms of the
the growth factor, $g(a):=\delta/a$, Eq.~(\ref{delta_gen}) can be written as
\begin{eqnarray}
&&\frac{\D^2 g}{\D a^2}+\frac{1}{a}\left(5+\frac{\D\ln H}{\D \ln a}\right)\frac{\D g}{\D a}
\nonumber\\&&\quad
+\frac{1}{a^2}\left[3+\frac{\D\ln H}{\D\ln a}-\frac{3}{2}\xi(a)\omm(a)\right]g=0.
\label{gr_fac}
\end{eqnarray}
We now introduce the new variable defined by\footnote{If the scale
factor is a multiple-valued function of $y$, it is not appropriate to use $y$
as a global time variable. This is indeed the case for the present Galileon model.
Even so,
the scale factor is a one-valued function of $y$ for sufficiently small $a$.}
\begin{eqnarray}
y:=\frac{1}{\omm(a)}-1.
\end{eqnarray}
In another way, it is written as $y=\rho_{\rm eff}/\rho_{\rm m}$.
In terms of $y$ Eq.~(\ref{gr_fac}) can be written as
\begin{eqnarray}
&&y(1+y)\frac{\D^2 g}{\D y^2}+\left[1+\frac{5}{2\zeta }+\left(\frac{3}{2}+\frac{5}{2\zeta }
\right)y
+\tilde\zeta\right]\frac{\D g}{\D y}
\nonumber\\&&
+\left[\frac{3+\zeta }{2\zeta^2 }-\frac{3}{2\zeta^2 }\frac{\xi(y)-1}{y}\right]g=0,
\label{gy}
\end{eqnarray}
where $\zeta:=\D\ln y/\D\ln a$ and $\tilde\zeta:=[(1+y)/\zeta]\D\ln\zeta/\D\ln a$.

In order to focus on analytically tractable cases, we consider the regime
in which the following two assumptions are valid.
The first assumption is that $\zeta=$ const and hence $\tilde\zeta=0$.
If dark energy is a fluid whose equation of state is $w=$ const,
then $\zeta = -3w$ and the assumption is valid over the whole history of the
universe after matter domination.\footnote{In the $w=$ const fluid model of dark energy,
the evolution of $\delta$ will be governed by Eq.~(\ref{delta_gen})
provided that the effect of clustering of the dark fluid is negligible.}
If instead the accelerated expansion is cause by modification of gravity,
$\zeta$ is not constant in general, but
at early times ($y\ll 1$) it is often good to approximate $\zeta=$ const.

The second assumption is that
$\xi$ can be approximated by $\xi- 1 = Ay+...$.
We do not consider the case where the leading term is given by $y^n$ with $n<1$, because if so then
$\delta$ cannot have the desired early-time behavior $\delta\approx a$ (i.e., $g\approx 1$).
If the leading term is $y^n$ with $n>1$ or $\xi=1$ exactly,
we neglect the effect of this term and set $A=0$.
An interesting example for which $A\neq 0$ is the DGP braneworld model~\cite{LueSS, KM, Linder2}.

Within the range of the validity of the above two assumptions,
the solution to Eq.~(\ref{gy}) is given by
\begin{eqnarray}
g&=&{}_2F_1(\alpha_+,\alpha_-; 1+5/(2\zeta); -y)
\nonumber\\
&=&1-\frac{3+\zeta-3A}{\zeta(5+2\zeta)}y+...,\label{g-sol}
\end{eqnarray}
where
\begin{eqnarray}
\alpha_\pm:=\frac{1}{4}\left(1+\frac{5}{\zeta}\right)\pm
\frac{1}{4}\sqrt{
\left(1+\frac{1}{\zeta}\right)^2+\frac{24A}{\zeta^2}
}.
\end{eqnarray}
The expression in the second line of Eq.~(\ref{g-sol}) is sufficient for our purpose.
However, the expression in the first line
might be useful if one considers
the model with $\xi=1$ (i.e., $A=0$) and $\zeta=$ const over the whole history of the universe
after matter domination because the expression in terms of the hypergeometric function then gives
the exact solution~\cite{exact1, exact2, exact3, exact4}.

Now the asymptotic value of the growth index
can computed as $\gamma_\infty=-\ln(1+\zeta \D\ln g/\D\ln y)/\ln(1+y)|_{y\to 0}= -\zeta\D g/\D y|_{y\to 0}$.
Thus, we obtain~\cite{Linder2}
\begin{eqnarray}
\gamma_\infty = \frac{3+\zeta-3A}{5+2\zeta}.\label{formula_gamma}
\end{eqnarray}

\subsection{Growth of matter density perturbations}

We now turn back to Galileon modified gravity and
study the growth index of density perturbations.
The metric perturbations in the Newtonian gauge is written as
\begin{eqnarray}
\D s^2=-(1+2\Phi)\D t^2+a^2(1-2\Psi)\D \mathbf{x}^2.
\end{eqnarray}
The perturbed energy-momentum tensor of cold dark matter is given by
\begin{eqnarray}
\delta T_0^{\;0}=-\delta\rho_{\rm m},\quad
\delta T_i^{\;0}=\partial_i\delta q_{\rm m},
\quad \delta T_i^{\;j}=0,
\end{eqnarray}
and the perturbation of the Galileon field is described by
$\sigma\to\sigma(t)+\delta\sigma(t,\mathbf{x})$.
From the traceless part of the gravitational field equations
we immediately obtain
\begin{eqnarray}
\Psi-\Phi=\delta\sigma.\label{per_traceless}
\end{eqnarray}
This is in contrast to the relation one finds in general relativity, $\Phi=\Psi$.

We are going to investigate the subhorizon evolution of the comoving density perturbation,
$\delta:=(\delta\rho_{\rm m}-3H\delta q_{\rm m})/\rho_{\rm m}$,
by employing the quasi-static approximation.
The quasi-static approximation was first introduced in~\cite{QSGR}
in the context of quintessence cosmological models
and then was applied to various theories of modified gravity~\cite{Reconst, KM, QS}.
The approximation amounts to
$\nabla^2X/a^2\gg H^2 X\gtrsim H\dot X, \ddot X$, where $X=\Phi, \Psi, \delta\sigma$,
and $\nabla^2:=\delta^{ij}\partial_i\partial_j$.
Although the full cosmological perturbation equations are involved~\cite{KTS},
the quasi-static approximation simplifies the field equations to
\begin{eqnarray}
\frac{\nabla^2}{a^2}{\cal E}_{\sigma }&=&0,\label{per_1}
\\
\frac{\nabla^2}{a^2}{\cal E}_{g}&=&\frac{\rho_{\rm m}}{2\phi}\delta,\label{per_2}
\end{eqnarray}
where
\begin{eqnarray}
{\cal E}_\sigma &:=&2\Psi-\Phi+\omega\delta\sigma
\nonumber\\&&
+\lambda^2\left[
 \dot\sigma^2 \Phi+
2\left( \ddot\sigma + \dot\sigma^2 
+2H \dot\sigma \right) \delta\sigma \right],
\\
{\cal E}_g&:=&2\Psi-\left(1-\lambda^2 \dot\sigma^2 \right) \delta\sigma .
\end{eqnarray}
The first equation is derived from the Galileon equation of motion
and the second from the gravitational field equations.
Note that the above equations are independent of the detailed structure of
the coupling function $\lambda^2(\sigma)$ in the sense that they do not include the terms
$(\lambda^2)_{,\sigma}$ and $(\lambda^2)_{,\sigma\sigma}$.
Using Eqs.~(\ref{per_traceless})--(\ref{per_2}) we obtain the
modified Poisson equation,
\begin{eqnarray}
\frac{\nabla^2}{a^2}\Phi =\frac{\rho_{\rm m}}{2\mpl^2}e^{-\sigma }
\left[1+\frac{(1+\lambda^2\dot\sigma^2 )^2}{\cal F}\right]\delta,
\label{poisson}
\end{eqnarray}
where
\begin{eqnarray}
{\cal F}:=3+2\omega+\lambda^2\left[4  \ddot\sigma
+2 \dot\sigma^2
+8H \dot\sigma -\lambda^2  \dot\sigma^4 \right].\label{def-F}
\end{eqnarray}
The perturbed energy-momentum conservation equations, $\delta\left(\nabla_\nu T_\mu^{\;\nu}\right)=0$,
yield
\begin{eqnarray}
\ddot\delta+2H\dot\delta = \frac{\nabla^2}{a^2}\Phi.\label{conservation}
\end{eqnarray}
Equations~(\ref{poisson}) and~(\ref{conservation}) are combined to give
the evolution equation for the density perturbation $\delta$
in the Galileon models~\cite{CK, SK},
which is of the form~(\ref{delta_gen})
with the identification $\xi=e^{-\sigma}[1+(1+\lambda^2\dot\sigma^2)^2/{\cal F}]$.
We have $\xi-1\ll 1$ initially,
and then $\xi$ increases with time, leading eventually to $\xi-1={\cal O}(1)$ today.
This effect enhances the growth rate of density perturbations~\cite{SK, KTS}.

As we already have the formula~(\ref{formula_gamma}) it is easy to
compute the asymptotic growth index $\gamma_{\infty}$ analytically.
At early times we have $\lambda^2\dot\sigma^2\approx 1/3$,
$e^\sigma\approx 1 + 2\dot\sigma/3H$, and ${\cal F}\approx 8H/3\dot\sigma$.
Noting that $y=\rho_{\rm eff}/\rho_{\rm m}\approx -4\dot\sigma/3H$, we find
$\xi\approx 1 +{\cal O}(y^2)$
and $\D\ln y/\D\ln a\approx 3/2$. This corresponds to $\zeta=3/2$ and $A=0$
in the formula~(\ref{formula_gamma}).
We thus arrive at
\begin{eqnarray}
\gamma_\infty = \frac{9}{16}= 0.5625.
\end{eqnarray}
This is certainly
different from the $\Lambda$CDM value, $\gamma_{\infty,\Lambda{\rm CDM}}=6/11= 0.5454...$,
but unfortunately the difference is only 3 percent.
The value is also different from the DGP value, $\gamma_{\infty, {\rm DGP}} = 11/16$,
because of the different rate of time-variation in the effective gravitational coupling.

\begin{figure}[tb]
  \begin{center}
    \includegraphics[keepaspectratio=true,height=55mm]{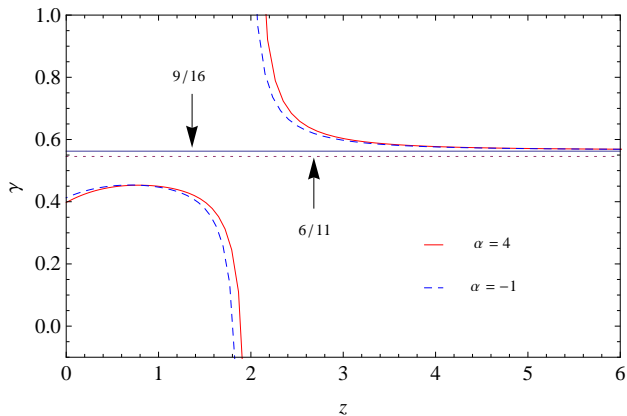}
  \end{center}
  \caption{$\gamma$ as a function of $z$ for $\alpha=4$ and $\alpha=-1$.
  The parameters are given by $\omega=-500$ and $\Omega_{{\rm m}0}=0.3$.}%
  \label{fig:gamma.eps}
\end{figure}

\begin{figure}[tb]
  \begin{center}
    \includegraphics[keepaspectratio=true,height=55mm]{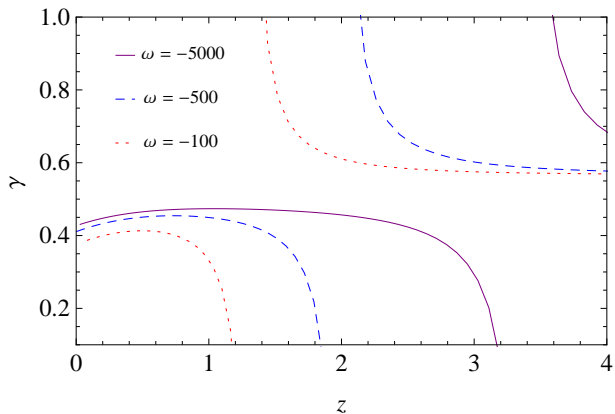}
  \end{center}
  \caption{$\gamma$ as a function of $z$ for $\alpha=1$ and various $\omega$.
  The parameters are given by $\Omega_{{\rm m}0}=0.3$.}%
  \label{fig:gamma2.eps}
\end{figure}

Numerical solutions to Eq.~(\ref{conservation}) with Eq.~(\ref{poisson}) are also easy to obtain.
Solving the background equations first as was done in the previous section and then
integrating the perturbation equation under the
initial condition $\delta\approx a$ at $t\approx 0$,
we have computed the growth index numerically.
The results are presented in Figs.~\ref{fig:gamma.eps} and~\ref{fig:gamma2.eps}.
One can confirm that the asymptotic value is indeed given by $\gamma_\infty = 9/16$.
Since $\omm (a)=1$ when $\rho_{\rm eff}=0$,
we have $\gamma\to\pm \infty$ as $z\to z_*^\pm$.
Similarly to the behavior of $w_{\rm eff}$ at $z=z_*$,
this is an artifact of the definition and the evolution of $\delta$ is not ill-behaved.
The detailed behavior at lower $z$ depends upon the parameters $\omega$ and $\alpha$,
but typically, $\gamma$ is as small as $\sim 0.4$ due to the enhanced growth of density perturbations.
Such a small value of $\gamma$ due to the enhanced growth of $\delta$ is also reported
in some $f(R)$ gravity models~\cite{gammafr}.\footnote{
The growth index in $f(R)$ gravity depends on the scale $k$,
in contrast to the present Galileon models.}
In any case, the growth index shows a distinct feature compared to the $\Lambda$CDM
model and other modified gravity models such as self-accelerating DGP braneworld.

\begin{figure}[tb]
  \begin{center}
    \includegraphics[keepaspectratio=true,height=55mm]{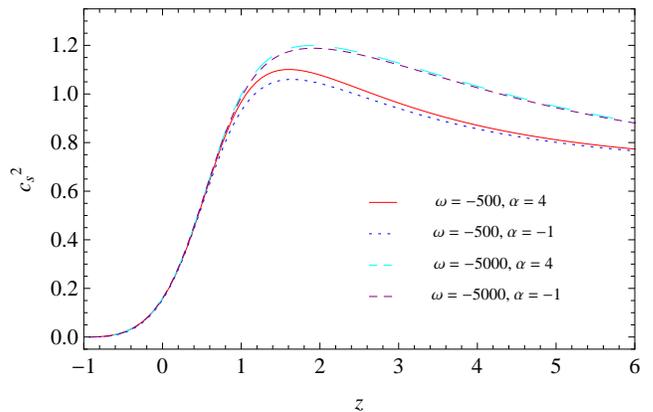}
  \end{center}
  \caption{The sound speed squared (defined in the Appendix) as a function of $z$.
  The sound speed squared is positive, and can be temporarily greater than 1.}%
  \label{fig:cs2.eps}
\end{figure}

\section{Conclusions}

In this paper, we have studied a class of Galileon cosmology,
i.e., cosmology in the Brans-Dicke theory
with the non-linear derivative interaction of the form $[\lambda^2(\phi)/\phi^2](\nabla\phi)^2\Box\phi$
with $\lambda^2(\phi)\propto \phi^{\alpha}$.
Thanks to this term in the Lagrangian, the Vainshtein mechanism works
and the scalar-mediated force is screened near matter sources.
The Vainshtein mechanism operates in the cosmological context as well,
leaving the cosmic expansion history indistinguishable from the usual matter-dominated universe.
Moreover, we have illustrated that
modification to gravity due to the Galileon field
can be responsible for the late-time acceleration of the universe,
closely mimicking the $\Lambda$CDM model or rather phantom dark energy models.
We have clarified that the late-time asymptotics depends on $\alpha$:
a future big rip singularity occurs for $\alpha<0$,
while
the phantom-like behavior is only temporal and we eventually have $w_{\rm eff}>-1$ for $\alpha>0$.
The result here is the generalization of the self-accelerating de Sitter universe found for $\alpha=0$
in~\cite{SK}.

We have also investigated the growth history of density perturbations in Galileon cosmology.
The asymptotic growth index
was obtained analytically as $\gamma_\infty=9/16$, which was confirmed numerically.
The growth index today was found to be $\gamma\sim 0.4$.
The behavior of the growth index is thus a clear discriminant
among Galileon scalar-tensor gravity, the $\Lambda$CDM model, and the DGP braneworld.


\acknowledgments
T.K. is supported by the JSPS under Contact Nos.~19-4199.

\appendix
\section{Stability analysis}

In the main text we advocate the ghost-free branch $\dot\sigma>0$.
The absence of ghost instabilities can be proved
by expanding the action at second order in the fluctuation of the Galileon field,
$\delta\sigma$.
Noting that the cubic term in the action is dominant at early times, we obtain
the following quadratic action~\cite{CK}:
\begin{eqnarray}
\delta S^{(2)}=6\int\D^4x\sqrt{-g}\,\phi\lambda^2 H\dot \sigma \left[
(\dot{\delta\sigma})^2-\frac{2}{3a^2}(\partial_i\delta\sigma)^2
\right].
\nonumber\\
\label{app-ac}
\end{eqnarray}
From this we see that $\dot \sigma>0$ is required in order to
avoid ghost instabilities. (The convention here is different from the one used in~\cite{CK}.
Our $\sigma$ is related to $\pi$ in~\cite{CK} as $\sigma\to -2\pi/\mpl$.)

Following Ref.~\cite{SK}
we neglect the friction term ($\propto\dot{\delta\sigma}$) and the effect of
metric perturbations on small scales to obtain the evolution equation for $\delta\sigma$:
\begin{eqnarray}
{\cal G}(t)\ddot{\delta\sigma}-{\cal F}(t)\frac{\nabla^2}{a^2}\delta\sigma=0,
\label{app-ev}
\end{eqnarray}
where ${\cal F}$ is defined in Eq.~(\ref{def-F}) and
\begin{eqnarray}
{\cal G}:=3+2\omega+\lambda^2\left(12H\dot\sigma+2\dot\sigma^2+3\lambda^2\dot\sigma^4\right)
-4(\lambda^2)_{,\sigma}\dot\sigma^2.\nonumber\\
\end{eqnarray}
(Full perturbation equations are found in the Appendix of Ref.~\cite{KTS}.)
It is easy to check that for $H^2\gg1/\lambda^{2}$
Eq.~(\ref{app-ev}) reduces to the equation of motion derived from the action~(\ref{app-ac}).
The condition for avoiding ghost instabilities is given by ${\cal G}>0$.
We also require $c_s^2:={\cal F}/{\cal G}>0$
so that the Galileon field is perturbatively stable.
Both are satisfied in our numerical examples presented in the main text
thanks to the non-linear term in the Lagrangian (see Fig~\ref{fig:cs2.eps}).
Note that the sound speed $c_s$ can be greater than the speed of light, but
that is a temporary behavior.
Superluminal propagation might indicate pathology and is claimed to be problematic
in Refs.~\cite{Adams:2006sv, Bonvin:2006vc, Ellis:2007ic, Bonvin:2007mw}.
However, some authors argue that propagation faster than light
does not necessarily lead to problems~\cite{Babichev:2007dw,Gorini:2007ta,Geroch:2010da}.
Thus, whether a theory with superluminal propagation is viable or not
is still a controversial issue.
We see that $c_s^2\to 0$ as $z\to -1$
and no particular behavior is found at the phantom divide crossing.



\begin{thebibliography}{99}


\bibitem{Copeland:2006wr}
For comprehensive review, see, e.g.,
  E.~J.~Copeland, M.~Sami and S.~Tsujikawa,
  Int.\ J.\ Mod.\ Phys.\  D {\bf 15}, 1753 (2006)
  [arXiv:hep-th/0603057].

\bibitem{f(R)}
For comprehensive review, see, e.g.,
T.~P.~Sotiriou and V.~Faraoni,
  arXiv:0805.1726 [gr-qc];
A.~De Felice and S.~Tsujikawa,
  arXiv:1002.4928 [gr-qc].



\bibitem{DGP}
G.~R.~Dvali, G.~Gabadadze and M.~Porrati,
  Phys.\ Lett.\  B {\bf 485}, 208 (2000)
  [arXiv:hep-th/0005016].

\bibitem{DDG}
C.~Deffayet, G.~R.~Dvali and G.~Gabadadze,
  Phys.\ Rev.\  D {\bf 65}, 044023 (2002)
  [arXiv:astro-ph/0105068].

\bibitem{will}
C.~M.~Will,
  Living Rev.\ Rel.\  {\bf 9}, 3 (2005)
  [arXiv:gr-qc/0510072].


\bibitem{uzan-review}
J.~P.~Uzan,
  arXiv:0908.2243 [astro-ph.CO].


\bibitem{chameleon}
D.~F.~Mota and J.~D.~Barrow,
  Phys.\ Lett.\  B {\bf 581}, 141 (2004)
  [arXiv:astro-ph/0306047];
  J.~Khoury and A.~Weltman,
  Phys.\ Rev.\ Lett.\  {\bf 93}, 171104 (2004)
  [arXiv:astro-ph/0309300];
 J.~Khoury and A.~Weltman,
  Phys.\ Rev.\  D {\bf 69}, 044026 (2004)
  [arXiv:astro-ph/0309411].


\bibitem{viablefr}
W.~Hu and I.~Sawicki,
  Phys.\ Rev.\  D {\bf 76}, 064004 (2007)
  [arXiv:0705.1158 [astro-ph]];
A.~A.~Starobinsky,
  JETP Lett.\  {\bf 86}, 157 (2007)
  [arXiv:0706.2041 [astro-ph]];
S.~A.~Appleby and R.~A.~Battye,
  Phys.\ Lett.\  B {\bf 654}, 7 (2007)
  [arXiv:0705.3199 [astro-ph]];
  S.~Nojiri and S.~D.~Odintsov,
  Phys.\ Rev.\  D {\bf 77}, 026007 (2008)
  [arXiv:0710.1738 [hep-th]].

\bibitem{Vs}
A.~I.~Vainshtein,
  Phys.\ Lett.\  B {\bf 39}, 393 (1972).

\bibitem{symmetron}
K.~Hinterbichler and J.~Khoury,
  arXiv:1001.4525 [hep-th].


\bibitem{G1}
A.~Nicolis, R.~Rattazzi and E.~Trincherini,
  Phys.\ Rev.\  D {\bf 79}, 064036 (2009)
  [arXiv:0811.2197 [hep-th]].

\bibitem{G2}
 C.~Deffayet, G.~Esposito-Farese and A.~Vikman,
  Phys.\ Rev.\  D {\bf 79}, 084003 (2009)
  [arXiv:0901.1314 [hep-th]];
  C.~Deffayet, S.~Deser and G.~Esposito-Farese,
  Phys.\ Rev.\  D {\bf 80}, 064015 (2009)
  [arXiv:0906.1967 [gr-qc]].

\bibitem{GE1}
C.~Charmousis, S.~C.~Davis and J.~F.~Dufaux,
  JHEP {\bf 0312}, 029 (2003)
  [arXiv:hep-th/0309083];
C.~Charmousis,
  Lect.\ Notes Phys.\  {\bf 769}, 299 (2009)
  [arXiv:0805.0568 [gr-qc]].


\bibitem{GE2}
E.~Elizalde, A.~G.~Zheksenaev, S.~D.~Odintsov and I.~L.~Shapiro,
  Class.\ Quant.\ Grav.\  {\bf 12}, 1385 (1995)
  [arXiv:hep-th/9412061].

\bibitem{Lovelock}
D.~Lovelock,
  J.\ Math.\ Phys.\  {\bf 12}, 498 (1971).


\bibitem{CK}
N.~Chow and J.~Khoury,
  Phys.\ Rev.\  D {\bf 80}, 024037 (2009)
  [arXiv:0905.1325 [hep-th]].

\bibitem{SK}
F.~P.~Silva and K.~Koyama,
  Phys.\ Rev.\  D {\bf 80}, 121301 (2009)
  [arXiv:0909.4538 [astro-ph.CO]].

\bibitem{KTS}
T.~Kobayashi, H.~Tashiro and D.~Suzuki,
  Phys.\ Rev.\  D {\bf 81}, 063513 (2010)
  [arXiv:0912.4641 [astro-ph.CO]].

\bibitem{extquint}
See, e.g.,
 J.~P.~Uzan,
  Phys.\ Rev.\  D {\bf 59}, 123510 (1999)
  [arXiv:gr-qc/9903004];
L.~Amendola,
  Phys.\ Rev.\  D {\bf 60}, 043501 (1999)
  [arXiv:astro-ph/9904120];
T.~Chiba,
  Phys.\ Rev.\  D {\bf 60}, 083508 (1999)
  [arXiv:gr-qc/9903094];
F.~Perrotta, C.~Baccigalupi and S.~Matarrese,
  Phys.\ Rev.\  D {\bf 61}, 023507 (2000)
  [arXiv:astro-ph/9906066];
O.~Bertolami and P.~J.~Martins,
  Phys.\ Rev.\  D {\bf 61}, 064007 (2000)
  [arXiv:gr-qc/9910056];
  G.~Esposito-Farese and D.~Polarski,
  Phys.\ Rev.\  D {\bf 63}, 063504 (2001)
  [arXiv:gr-qc/0009034];
  E.~Gunzig, A.~Saa, L.~Brenig, V.~Faraoni, T.~M.~Rocha Filho and A.~Figueiredo,
  Phys.\ Rev.\  D {\bf 63}, 067301 (2001)
  [arXiv:gr-qc/0012085];
  T.~Chiba,
  Phys.\ Rev.\  D {\bf 64}, 103503 (2001)
  [arXiv:astro-ph/0106550];
  D.~F.~Torres,
  Phys.\ Rev.\  D {\bf 66}, 043522 (2002)
  [arXiv:astro-ph/0204504];
  R.~Gannouji, D.~Polarski, A.~Ranquet and A.~A.~Starobinsky,
  JCAP {\bf 0609}, 016 (2006)
  [arXiv:astro-ph/0606287];
  S.~Carloni, S.~Capozziello, J.~A.~Leach and P.~K.~S.~Dunsby,
  Class.\ Quant.\ Grav.\  {\bf 25}, 035008 (2008)
  [arXiv:gr-qc/0701009];
  S.~Tsujikawa, K.~Uddin, S.~Mizuno, R.~Tavakol and J.~Yokoyama,
  Phys.\ Rev.\  D {\bf 77}, 103009 (2008)
  [arXiv:0803.1106 [astro-ph]];
T.~Chiba, M.~Siino and M.~Yamaguchi,
  arXiv:1002.2986 [astro-ph.CO].



\bibitem{Reconst}
 B.~Boisseau, G.~Esposito-Farese, D.~Polarski and A.~A.~Starobinsky,
  Phys.\ Rev.\ Lett.\  {\bf 85}, 2236 (2000)
  [arXiv:gr-qc/0001066].


\bibitem{Peeb}
P.~J.~E.~Peebles,
{\em The Large-Scale Structure of the Universe}
(Princeton University Press, Princeton, New Jersey, 1980);
P.~J.~E.~Peebles,
  Astrophys.\ J.\  {\bf 284}, 439 (1984).

\bibitem{WS}
  L.~M.~Wang and P.~J.~Steinhardt,
  Astrophys.\ J.\  {\bf 508}, 483 (1998)
  [arXiv:astro-ph/9804015].

\bibitem{Linder1}
E.~V.~Linder,
  Phys.\ Rev.\  D {\bf 72}, 043529 (2005)
  [arXiv:astro-ph/0507263].
  
\bibitem{Linder2}
E.~V.~Linder and R.~N.~Cahn,
  Astropart.\ Phys.\  {\bf 28}, 481 (2007)
  [arXiv:astro-ph/0701317].

\bibitem{P-G}
D.~Polarski and R.~Gannouji,
  Phys.\ Lett.\  B {\bf 660}, 439 (2008)
  [arXiv:0710.1510 [astro-ph]].

\bibitem{SS}
V.~Sahni and Y.~Shtanov,
  JCAP {\bf 0311}, 014 (2003)
  [arXiv:astro-ph/0202346].

\bibitem{ndgp1}
A.~Lue and G.~D.~Starkman,
  Phys.\ Rev.\  D {\bf 70}, 101501 (2004)
  [arXiv:astro-ph/0408246].


\bibitem{ndgp2}
L.~P.~Chimento, R.~Lazkoz, R.~Maartens and I.~Quiros,
  JCAP {\bf 0609}, 004 (2006)
  [arXiv:astro-ph/0605450].

\bibitem{LueSS}
A.~Lue, R.~Scoccimarro and G.~D.~Starkman,
  Phys.\ Rev.\  D {\bf 69}, 124015 (2004)
  [arXiv:astro-ph/0401515].

\bibitem{KM}
K.~Koyama and R.~Maartens,
  JCAP {\bf 0601}, 016 (2006)
  [arXiv:astro-ph/0511634].


\bibitem{exact1}
D.~J.~Heath, Mon.\ Not.\ R.\ Astron.\ Soc.\ {\bf 179}, 351 (1977).
\bibitem{exact2}
S.~Bildhauer, T.~Buchert and M.~Kasai, Astron.\ Astrophys.\ {\bf 263}, 23 (1992).
\bibitem{exact3}
V.~Sahni and P.~Coles,
  Phys.\ Rept.\  {\bf 262}, 1 (1995)
  [arXiv:astro-ph/9505005].
\bibitem{exact4}
S.~Lee and K.~W.~Ng,
  arXiv:0906.1643 [astro-ph.CO].

\bibitem{QSGR}
A.~A.~Starobinsky,
  JETP Lett.\  {\bf 68}, 757 (1998)
  [Pisma Zh.\ Eksp.\ Teor.\ Fiz.\  {\bf 68}, 721 (1998)]
  [arXiv:astro-ph/9810431].



\bibitem{QS}
 S.~Tsujikawa,
  Phys.\ Rev.\  D {\bf 76}, 023514 (2007)
  [arXiv:0705.1032 [astro-ph]];
S.~Tsujikawa, K.~Uddin and R.~Tavakol,
  Phys.\ Rev.\  D {\bf 77}, 043007 (2008)
  [arXiv:0712.0082 [astro-ph]].

\bibitem{gammafr}
R.~Gannouji, B.~Moraes and D.~Polarski,
  JCAP {\bf 0902}, 034 (2009)
  [arXiv:0809.3374 [astro-ph]];
  S.~Tsujikawa, R.~Gannouji, B.~Moraes and D.~Polarski,
  Phys.\ Rev.\  D {\bf 80}, 084044 (2009)
  [arXiv:0908.2669 [astro-ph.CO]];
T.~Narikawa and K.~Yamamoto,
  arXiv:0912.1445 [astro-ph.CO];
  H.~Motohashi, A.~A.~Starobinsky and J.~Yokoyama,
  arXiv:1002.1141 [astro-ph.CO].


\bibitem{Adams:2006sv}
  A.~Adams, N.~Arkani-Hamed, S.~Dubovsky, A.~Nicolis and R.~Rattazzi,
  JHEP {\bf 0610}, 014 (2006)
  [arXiv:hep-th/0602178].

\bibitem{Bonvin:2006vc}
  C.~Bonvin, C.~Caprini and R.~Durrer,
  Phys.\ Rev.\ Lett.\  {\bf 97}, 081303 (2006)
  [arXiv:astro-ph/0606584].

\bibitem{Ellis:2007ic}
  G.~Ellis, R.~Maartens and M.~A.~H.~MacCallum,
  Gen.\ Rel.\ Grav.\  {\bf 39}, 1651 (2007)
  [arXiv:gr-qc/0703121].

\bibitem{Bonvin:2007mw}
  C.~Bonvin, C.~Caprini and R.~Durrer,
  arXiv:0706.1538 [astro-ph].

  
\bibitem{Babichev:2007dw}
  E.~Babichev, V.~Mukhanov and A.~Vikman,
  JHEP {\bf 0802}, 101 (2008)
  [arXiv:0708.0561 [hep-th]].

\bibitem{Gorini:2007ta}
  V.~Gorini, A.~Y.~Kamenshchik, U.~Moschella, O.~F.~Piattella and A.~A.~Starobinsky,
  JCAP {\bf 0802}, 016 (2008)
  [arXiv:0711.4242 [astro-ph]].

\bibitem{Geroch:2010da}
  R.~Geroch,
  arXiv:1005.1614 [gr-qc].

\end{thebibliography}
\end{document}